\documentclass[pss]{wiley2sp} % provides new 2008 pss two-column style (no alternative manuscript style output available at present)
\usepackage{amsmath}
\usepackage{bm}
\usepackage{w-greek}
\usepackage{amssymb}

\begin{document}

% Title of the article
\title{Charge Density Wave in Sr$_{14-x}$Ca$_x$Cu$_{24}$O$_{41}$}

% Abbreviated title for the page headers
\titlerunning{Charge Density Wave...}

% Authors
\author{%
  Krzysztof Wohlfeld\textsuperscript{\Ast,\textsf{\bfseries 1}},
  Andrzej M. Ole\'s\textsuperscript{\textsf{\bfseries 1}},
  George A. Sawatzky\textsuperscript{\textsf{\bfseries 2}}}

% Abbreviated list of authors for the page headers
\authorrunning{Krzysztof Wohlfeld et al.}

%E-mail-address of corresponding author
\mail{e-mail
  \textsf{k.wohlfeld@ifw-dresden.de}}

% author's affiliations/addresses
\institute{%
  \textsuperscript{1}\,Marian Smoluchowski Institute of Physics, Jagellonian
 University, Reymonta 4, PL-30059 Krak\'ow, Poland\\
  \textsuperscript{2}\,Department of Physics and Astronomy, University of British
Columbia, Vancouver B. C. V6T-1Z1, Canada }

%\received{XXXX, revised XXXX, accepted XXXX} % do not change, will be filled in by the publisher
\published{: Phys. Stat. Sol. (b) 247, 668 (2010)} % do not change, will be filled in by the publisher

%Please select four to six PACS-codes from the enclosed list (PACS.txt) or from www.aip.org/pacs)
 % For example: 71.20.Ps

\abstract{%
% This is a macro for the typesetting of two-column text in an
% abstract. It will typeset the two arguments in \abstcol{}{} as the
% left and right column inside the abstract box. At the
% columnbreak there will be always a columnbreak (\par), so both
% columns start with a new paragraph. No automatic column height
% balancing is done.
%
% If used with a \titlefigure it will silently output both
% parameters as consecutive paragraphs.
%
% The macro is defined exclusively inside the argument of \abstract{};
% if used outside it will raise an error.
%
% Usage: \abstcol{<left column>}{<right column>}
\abstcol{%
Recent experimental results \cite{Rus06,Rus07} revealed the
stability of a peculiar charge density wave (CDW) with odd period
($\lambda=3,5$) in the coupled Cu$_2$O$_5$ spin ladders in
Sr$_{14-x}$Ca$_x$Cu$_{24}$O$_{41}$, while the CDW with
even period is unstable.
Here we resolve this contradiction by solving the effective
$t$--$J$ model for the two coupled chains {\it supplemented} by
the intersite interaction term $V$.
  }{%
Crucial to this finding is the novel interaction term $V$
(responsible for the interladder repulsion between nearest
neighbor holes), which is derived from the on-site Hubbard-like
repulsion between electrons on the same oxygen sites but within
two distinct orbitals which belong to two neighboring Cu$_2$O$_5$
ladders in the charge transfer model \cite{Woh07}. }}

% The class file requires the standard graphicx Latex package. See the 'LaTeX
% standard graphics and color packages documentation' for more information at
% <http://tug.ctan.org/tex-archive/macros/latex/required/graphics/grfguide.pdf>.
%
% Accepted figure file formats depend on which LaTeX flavour is used.
% Classic LaTeX is always able to use Encapsulted Postscript (EPS);
% PDFLaTeX can't use this but accepts PDF, JPG, PNG, and GIF formats.
%
% See examples for implementing graphics in floating figure environments later in this file.
% If \titlefigure is given, it takes as its mandatory parameter the
% name (without extension) of some figure file

\maketitle   % please do not remove

\section{Introduction}
It is at the heart of the understanding of the high-temperature
superconducting (SC) phase to infer what kind of state competes
with the SC one in the possible quantum phase transition between
the two. As in the two-dimensional (2D) copper oxides one of the
candidates of such a state is the stripe phase, it is natural to
expect that in the quasi-2D copper oxide coupled-ladder system
also a kind of charge ordered state would compete with the
predicted there SC state \cite{Dag92}. Indeed in
Sr$_{14-x}$Ca$_x$Cu$_{24}$O$_{41}$, which has the SC state for
$x=13.6$ and under pressure larger than 3 GPa \cite{Ueh96}, a
charge density wave (CDW) phase with period $\lambda=3$ and
$\lambda=5$ was observed when the Calcium concentration was tuned
to $x=11$ (which corresponds to the hole concentration $n_h=6/5$
per copper) and $x=0$ (where $n_h=4/3$), respectively
\cite{Rus06,Rus07}.
Although the charge transfer model 
for coupled ladders proposed in Ref. \cite{Woh07} may explain the onset 
of such an odd-period CDW state, it is the $t$-$J$-like model which is 
{\it a natural} Hamiltonian for the doped {\it spin} ladders \cite{Rus06,Rus07}.
However, then this experimental observation is a challenge for
the theory since such a novel CDW state with the odd period has
not been predicted by the standard $t$--$J$ model for the 
ladders \cite{Whi02}. 
Moreover, the standard $t$--$J$ model
predicts the CDW order with period $\lambda=4$ to be stable for
$n_h=5/4$ which would correspond to the $x=4$ doping in
Sr$_{14-x}$Ca$_x$Cu$_{24}$O$_{41}$ where no charge order was
observed \cite{Rus06,Rus07}.

The purpose of this paper is to shed some light on this apparent
contradiction between the experiment and the theory. 
Therefore, first we discuss which terms are missing in the standard
$t$--$J$ model while in the second part of the paper
we look at the generic features of this new model by solving it in
a simplified case for two coupled chains using the Gutzwiller
approach and the mean-field approximation.

\section{Model}
Our starting point is the effective one-dimensional (1D) $t$--$J$
model for a single chain,
\begin{align}\label{eq:1}
H&=-t\,\sum_{i \sigma } \Big(\tilde{d}^{\dag}_{i \sigma}
\tilde{d}_{i+1,\sigma}^{}+ {\rm H.c.}\Big)
\nonumber \\
&+ J \sum_{i} \Big({\bf S}_{i} \cdot {\bf S}_{i+1} -
\frac{1}{4}\,\tilde{n}_{i d} \tilde{n}_{i+1, d} \Big)\,,
\end{align}
where: ($i$) the operator $\tilde{d}^\dag_{i\sigma}=d^\dag_{i
\sigma}(1-n_{id,-\sigma})$ creates an electron at site $i$ with
spin $\sigma$ in the restricted Hilbert space with no double
occupancies, ($ii$) $\tilde{n}_{id\sigma}=
\tilde{d}^\dag_{i\sigma}\tilde{d}_{i\sigma}^{}$ is the
corresponding electron number operator, and ($iii$) ${\bf S}_i$ is
an $S=1/2$ spin operator. Besides, the average number of $d$ holes
per site here is $n=2-n_h$, with
$n=\sum_{\sigma}\langle\tilde{n}_{id\sigma}\rangle$.

\begin{figure}[t]%
\begin{center}
\includegraphics*[width=0.9\linewidth]{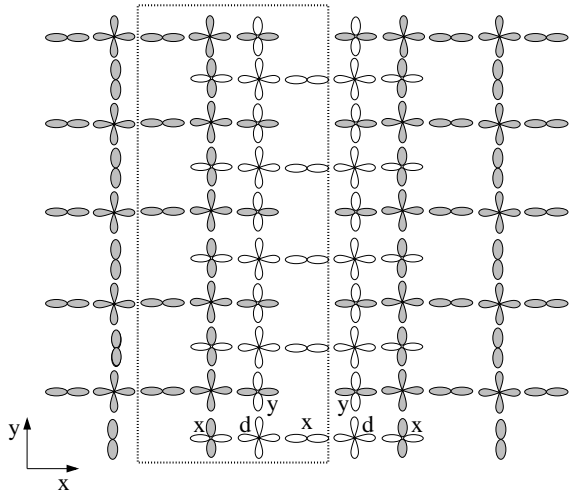}
\end{center}
\caption{ Artist's view of the three coupled Cu$_2$O$_5$ ladders
(denoted by grey, white, and again grey color, respectively). Only
orbitals which are included in the charge transfer model of Ref.
\cite{Woh07} are shown (where $x\equiv 2p_x$, $y\equiv 2p_y$, and
$d\equiv 3d_{x^2-y^2}$). The vertical box depicts two coupled
CuO$_3$ chains belonging to two different neighboring ladders and
discussed in the paper.}
\label{fig:1}
\end{figure}

In the standard approach to the Cu$_2$O$_5$ systems this model is
derived from the charge transfer model which is similar to that
for CuO$_2$ planes \cite{Ole87}. Then the model should also
contain the kinetic $\propto t$ and superexchange $\propto J$
terms along the rungs of the ladder
--- it would then constitute the standard $t$--$J$ model for the
ladder \cite{Whi02}. However, a closer look at the geometry of the
Cu$_2$O$_5$ coupled ladders reveals that even such a model would
not be enough. In fact, one cannot neglect a Coulomb repulsion
on-site element between two holes on the same oxygen sites
($\propto U_p\gg t_{pd}$ \cite{Gra92}, where $t_{pd}$ is the Cu--O
hybridization) but on two different oxygen $2p$ orbitals ($x$
and $y$ orbitals in Fig. \ref{fig:1}) belonging to the two
neighboring chains in different ladders \cite{Woh07}. Thus,
although the holes do not hop between the neighboring Cu$_2$O$_5$
ladders (a small inter-oxygen hopping can be neglected
\cite{Gra92}) the $t$--$J$ model for the single ladder should
contain an extra {\it interladder} coupling which can be derived
from this Coulomb interaction.

Altogether this means that neglecting the kinetic and
superexchange terms along the rung connecting the two chains and
taking into account the interaction between the two neighboring
chains is {\it a priori} as justified in the treatment of
Cu$_2$O$_5$ coupled ladders as skipping the interladder coupling and
considering merely the $t$--$J$ model for a single ladder.
Naturally, the proper approach is to take into account both of
these effects (under which the work is in progress) but in what
follows we merely concentrate on the problem of two coupled chains
connected by this interladder coupling, see Fig. \ref{fig:1}. This
will enable us to study the generic role of the interladder
coupling.

\begin{figure}[t]%
\begin{center}
\includegraphics*[width=0.75\linewidth]{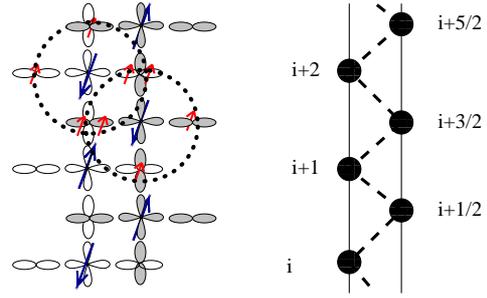}
\end{center}
\caption{%
Left panel: the artist's view of the interladder repulsion between
two nearest neighbor Zhang-Rice singlets on two nearest neighbor
chains. The Zhang-Rice state is depicted by a dotted ring. Large
(small) arrows depict the hole spins for +1.0 (+0.25) charge.
Right panel: the geometry of the model described by the $H+\bar{H}+H'$
Hamiltonian, see text for further details.} \label{fig:2}
\end{figure}

While the detailed derivation of the interladder coupling is
somewhat lengthy and will be described elsewhere, a simple
cartoon-picture shown in the left panel of Fig. \ref{fig:2}
explains its basic idea. Here we see that the two Zhang-Rice
singlets \cite{Zha88} which are situated next to each other in the
two neighboring chains (see dotted rings in Fig. \ref{fig:2})
share the same oxygen sites. More precisely, as an oxygen hole
forming a Zhang-Rice singlet state is equally distributed among
the four oxygen sites surrounding the central copper hole
\cite{Zha88}, there is a $0.25$ hole charge in each of these
oxygen orbital sharing the common oxygen sites. Since the two holes
in two different orbitals repel as $\propto U_p$ (where the Hund's
exchange $J_H$ is skipped for simplicity), the Zhang-Rice singlets
repel as $\propto (1/4 \times 1/4 + 1/4 \times 1/4 )U_p$.
Obviously, the Zhang-Rice states in each chain are not the proper
orthogonal states (similarly as in the 2D case \cite{Zha88}) and
one needs to orthogonalize them first. This, however, does not
change the result qualitatively as it occurs that such a procedure
yields that the two Zhang-Rice singlets repel as $\propto 0.14
U_p$. Besides, finite Hund's exchange reduces this interaction by
a factor $(1-5J_H/2U_p)$.

Hence, we supplement the $t$--$J$ model for a single chain, Eq.
(\ref{eq:1}), by the following interladder term which couples the
two chains under consideration
\begin{align}\label{eq:2}
H' = V \sum_{i}\Big( \tilde{n}_{i d} \tilde{\bar{n}}_{i+\frac12,
d} + \tilde{n}_{i d} \tilde{\bar{n}}_{i-\frac12, d} \Big),
\end{align}
where $V=0.14 U_p(1-5J_H/2U_p)$ and for realistic parameters
\cite{Woh07} $V\sim 0.5t$. Here the {\it bar} signs are added over
the operators which act in the Hilbert subspace of the neighboring
chain. The detailed geometry of the full model for two coupled
chains ${\cal H}\equiv
H+\bar{H}+H'$ can be seen in the right panel of Fig. \ref{fig:2} --- the
solid lines show the bonds along which the kinetic and
superexchange term of Eq. (\ref{eq:1}) act, while the dashed lines
show the bonds along which the interladder terms of Eq.
(\ref{eq:2}) are finite.

\section{Results}
In order to solve the $t$--$J$--$V$ model, given by the Hamiltonian
${\cal H}$, we first need to overcome the constraint
of no double occupancies. One of the approximate ways to do it is
to introduce the Gutzwiller factors $g_t=(2-2n)/(2-n)$ and
$g_J=4/(2-n)^2$, which renormalize the kinetic term ($g_t$) and the
superexchange and interladder terms ($g_J$). Similar factors were used recently to
investigate the superconducting flux phases in the cuprates
\cite{Rac07}. Then the whole Hamiltonian is defined in terms of
unrestricted fermions $d_{i\sigma}^{\dagger}$.

Next, we decouple the fermion operators in the superexchange and
interladder terms in a mean-field way:
$d^{\dag}_{i\sigma}d_{i\sigma}d^{\dag}_{i\sigma'}d_{i\sigma' }
\rightarrow d^{\dag}_{i\sigma}d_{i\sigma }\langle
d^{\dag}_{i\sigma'}d_{i\sigma'}\rangle +\langle
d^{\dag}_{i\sigma}d_{i\sigma}\rangle
d^{\dag}_{i\sigma'}d_{i\sigma'}$ (a similar decoupling is used
also for the {\it bar} operators). Then, we diagonalize the
effective one-particle Hamiltonian in $k$ space assuming that the
classical fields take their initial values as
\begin{equation}\label{eq3:l5}
\langle d^{\dag}_{i\sigma}d_{i\sigma}\rangle =\left\{
\begin{array}{cc}
n-p & {\rm for}\  i/\lambda \in \mathbb{Z} \\
n+\frac{1}{\lambda-1} p & {\rm for }\ i/\lambda \notin\mathbb{Z}
\end{array} \right. \ ,
\end{equation}
which defines the CDW order parameter $p$.
As already mentioned, here $\lambda$ is the CDW period, 
cf. Fig. 1 in Ref. \cite{Woh07}.
In addition,
$\langle\bar{d}^{\dag}_{i-\frac12, \sigma}\bar{d}_{i-\frac12 \sigma}\rangle$ are
defined as in Eq. (\ref{eq3:l5}) but with $i/\lambda$ replaced by 
$(i+1)/\lambda$ [$(i+2)/\lambda$ for period $\lambda=5$] -- such assumption
minimizes the classical energy cost of the interladder repulsion $V$.
Finally, the actual values of the classical fields (and
consequently the order parameter $p$) are obtained by iterating
the above procedure (i.e. calculated from the diagonalized
Hamiltonian) until their initial and final values converge.

\begin{figure}[t]%
\begin{center}
\includegraphics*[width=0.86\linewidth]{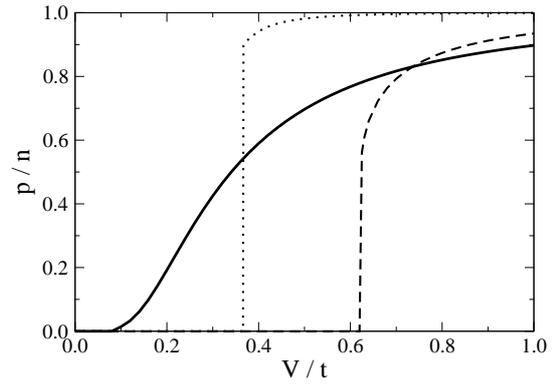}
\end{center}
\caption{The CDW order parameter $p$ as a function of the
interladder interaction $V$ in the two coupled chains with
$J=0.4t$: for filling $n=2/3$ ($n_h=4/3$) with period $\lambda=3$
(solid line), for filling $n=3/4$ ($n_h=5/4$) with period
$\lambda=4$ (dashed line), and for filling $n=4/5$ ($n_h=6/5$) with
period $\lambda=5$ (dotted line).} \label{fig:3}
\end{figure}

As the main result of this self-consistent procedure, we obtained
the increasing order parameter $p$ as a function of the
interladder interaction $V$ for a realistic value of $J=0.4t$
\cite{Gra92} and for the three interesting hole dopings, see
Fig. \ref{fig:3}. First, we see that for finite values of $V$ the
CDW phase with the peculiar odd period is stable. Moreover, it is
stable for realistic values of parameter $V\sim 0.5t$. Second, the
CDW order with period $\lambda=4$ is less stable as the CDW order
sets in only for $V>0.6t$.

One may ask, why the CDW phase with even period appears to be less
stable. Although this reproduces the experimental result quite
well, it is a very peculiar phenomenon associated purely with the
1D physics. In fact, it is the CDW phase with period $\lambda=3$
which is remarkably stable here due to the nesting of the band
structure for this particular doping, and indeed we find the
exponential growth of $p$ with increasing $V>0$, special for the
present 1D model.

\section{Conclusions}
In summary, we showed that including the Coulomb repulsion between
holes on the same oxygen sites but in different orbitals and
belonging to two different neighboring Cu$_2$O$_5$ ladders, led to
the effective repulsion between Zhang-Rice singlets which in turn
led to the interladder repulsion $V$ between nearest neighbor
holes situated in the neighboring ladders. Next, we solved a
$t$--$J$ model supplemented by the repulsive term $V$ in a
simplified geometry of two coupled chains. It occurred that the
CDW state with odd period was stable in such a model for the
realistic values of the interaction parameters $V$, while the CDW
state with the even period ($\lambda=4$) was stable merely for
slightly enhanced values of these parameters.

The results presented in this paper could possibly explain the
onset of the novel odd-period-CDW state in the Cu$_2$O$_5$ coupled
ladder system in Sr$_{14-x}$Ca$_x$Cu$_{24}$O$_{41}$ with $x=0$ and
$x=11$. However, a detailed study of this phenomenon is currently
under investigation since one needs to generalize the
$t$--$J$--$V$ model presented here to the coupled ladders and 
to investigate the stability of CDW states in this generalized 
model in order to verify this conjecture.

\begin{acknowledgement}
We wish to acknowledge financial support by the Foundation for
Polish Science (FNP) and the Polish Ministry of Science and Higher
Education under Project No.~N202 068 32/1481.
\end{acknowledgement}

% Use the following code if you wish to generate your bibliography with BibTeX;
% replace the string "pss-demo" below with the name(s) of
% the BibTeX data base(s) you want to use.
% The resulting bibliography-output (the contents of the .bbl file)
% must be pasted back into this file before submission.
%
% \bibliographystyle{pss}
% \bibliography{pss-demo}
%
% Replace the following example bibliography with your references
% before submission:

\end{document}